\documentclass[aps,twocolumn]{revtex4} 
\usepackage{amsmath}
\usepackage[dvips]{graphicx}
\usepackage{graphics,epsfig}

\newcommand{\be}{\begin{equation}}
\newcommand{\ee}{\end{equation}}
\newcommand{\ba}{\begin{eqnarray}}
\newcommand{\ea}{\end{eqnarray}}
\newcommand{\lb}{\label}

\newcommand{\half}{\frac{1}{2}}

\newcommand{\nn}{\nonumber}
\begin{document}
\title{Simple pairs potential-density for  flat rings}
\author{Patricio  S. 
Letelier}
\email{ letelier@ime.unicamp.br}
 \affiliation{Departamento de Matem\'{a}tica Aplicada\\ Instituto de Matem\'{a}tica, Estat\'\i stica e Computa\c c\~ao Cient\'\i fica\\ Universidade Estadual
de Campinas \\13083-970 Campinas, S.\ P., Brazil}

\begin{abstract}
Pairs potential-density in terms of elementary functions that represents
 flat rings structures are presented. We study structures representing one or
 several concentric flat rings. Also disks surrounded by concentric flat rings are 
exhibited. The stability of concentrically circular orbits of particles moving on a 
flat ring structure is analyzed for  radial perturbations.
\end{abstract}
\maketitle
\section{Introduction}
Flat rings systems are a common feature of the four giant planets of the 
Solar System.  Also ringy structures are  present in several observed galaxies and nebulae, e.g.,
in the Hourglass nebula we have as an outstanding  characteristic: the presence of a very net and large isolated ring. The exact gravitational potential  of a ring of zero
thickness and constant linear density is given in its exact form by an elliptic 
integral that is seldom used for practical purposes. This  potential is usually approximated by  truncated series of spherical harmonics, i.e., a multipolar expansion. As far as we know 
there not
simple expressions for  the exact gravitational potential of a  flat ring of any surface density.

The potential of  a  ring enclosing a disk can be obtained by a process of complexification 
of the potential  of a punctual mass \cite{appell}, \cite{WW}, \cite{gleisser}. This potential con be used to built a family of
similar structures \cite{letol}. On the other hand simple pairs potential-density for thin disks are
known. The simpler  is the Plummer-Kuzmin disk [\cite{plummer} and \cite{kuzmin}] that represents a simple model of galactic disk 
with a concentration of mass in its center and density that decays as $1/r^3$ on the plane of the disk, see also \cite{BT}. This structure has 
no boundary even though for practical purposes one can put a cutoff radius  wherein the main
part of the mass is inside, say 98\% of the mass. Another simpler models of disks are the \cite{MM} disks. These  disks have a mass concentration on their centers and finite radius.
The \cite{lord} inversion theorem can be used to invert the Morgan and Morgan disks to 
produce an infinite disk with a central hole of the same radius of the original disk.
We can also put a cutoff in these inverted disks. Therefore the inverted Morgan and Morgan disks can be considered as representing a flat rings. In the context Einstein theory of gravitation  one of this  inverted rings was used to study the gravity of a disk with a  central black hole, \cite{lemlet}.

The purpose of this article is to use the family of Morgan and Morgan disks to construct
flat ring like structures. We shall use two different approaches, the first consists in  the
superposition of finite disks of different densities. By using  this method we can construct structures that represent one or several concentric flat rings and families of   disks surrounded by flat rings. These structures  have a finite outer radius. The second method is the Lord  Kelvin inversion method that will be used to
construct the corresponding  inverse structures obtained using the first method. In this case the structures  extend to infinite, but as before  one can also put a cutoff. 

 As a very first test of stability for these structures we study the linear stability of particles moving in circular orbits inside them. In other words, we assume the very naive model that the structures are build from particles moving only in concentric  circles. The stability of circular orbits in several axially symmetric  systems is studied in both Newton and Einstein theories of gravitation in \cite{orbits}.

This article is divided as follows, in Sec. II we give a quick revision of the pair potential-density associated to the Morgan and Morgan disks. In Sec. III we study several
classes of superpositions that give rise to pairs potential-density of structures that can be considered as representing  one or several concentric flat rings. In the next section, Sec. IV, we construct pairs 
potential-density associated to structures representing a central disk surrounded by one or several concentric flat rings. In the following
section, Sec. V, we study the Kelvin inversion of the previous studied structures. In the penultimate 
section, Sec VI, we study the stability under linear radial perturbations of four of the  ring structures already presented. Finally, in Sec VII, we summarize and discuss some of the previous results.
\section{Morgan and Morgan disks}
The \cite{MM} disks are obtained by solving the Laplace equation in  the natural coordinates to represent the gravitational potential of a disk like structure, i.e., prolate coordinates $(\nu, \mu, \varphi)$
that are related to the usual cylindrical coordinates $(r,z,\varphi)$ by
\be
 a\nu= -{\rm Im} \mathcal{ R},  \;\;  a\mu= {\rm Re}{\mathcal R},\;\;
  \mathcal{R}\equiv \sqrt{r^2+(z-ia)^2},
\lb{prolate1}
\ee
where $a$ is a positive constant.
From the previous equations we find
\be
r=a\sqrt{\nu^2+1)(1-\mu^2)}, \;\; z=a\nu\mu.   \lb{prolate2}
\ee
Note that on the plane  $z=0$  we have  $\nu=0$ and $\mu=\sqrt{1-p^2}$, with $p=r/a$.

From the Laplace equation in prolate coordinates one can find, \cite{bateman}, the potential
of a disk parallel to the plane $z=0$ and centered on the origin of the coordinate system,
\be
V_{2m}=(1-2m)GM_{2m}(-1)^m P_{2m}(0)P_{2m}(\mu) q_{2m}(\nu)/a, \lb{V2m}
\ee
where $M_{2m}$ is the mass of the disk with potential $V_{2m}$ and $a$ its radius.
$P_{n}(x)$ are the usual Legendre polynomials and $q_n(x)$  are related to the usual
Legendre functions, $Q_n(x)$,  by $ q_n(x)=i^{n+1} Q_n(i x) $.  We recall the identity,
\ba
(-1)^n P_{2n}(0)&=&1\cdot 3\cdot 5 \; ... \; (2n-1)/(2 \cdot 4 \cdot 6 \; ...\; 2n),  \nonumber\\
&=&(2n-1)!!/(2n)!!,
\ea
that will  frequently used.

Associated to the potential (\ref{V2m}) we have the  surface density, \cite{bateman},
\be
\sigma_{2m}=(2m+1)M_{2m}P_{2m}(1-p^2)/(2\pi a^2\sqrt{1-p^2}).
\lb{density1}
\ee
\cite{MM} considered the following superposition of the above mentioned solutions of Laplace equation,
\be
V^{(n)}=\sum_{m=0}^{m=n}A_{2m,2n}V_{2m} \lb{V2n},
\ee
with
\be
A_{2m,2n}=\frac{(4m+1)2^{2m}(2n)!}{(n+m)!(n-m)!(2n+2m+1)!}.
\lb{Amn}
\ee
This superposition represents the potential of disks with the simple associated surface density,  
\be
\sigma^{(n)}(r)=(2n+1)M^{(n)}(1-p^2)^{n-\half}/(2\pi a^2),
\lb{density2}
\ee
where we have taken  $M_{2n}=M^{(n)}$  for all $n$.

The potential on the plane $z=0$ can be obtained using the algorithm, \cite{MM},
\be
V^{(n)}=-\frac{GM^{(n)}}{a}(2n+1)(-1)^n P_{2n}(0)\int_{0}^{\pi/2}[1-p^2\cos^{2} \theta]^n d\theta
\ee
The $n=0$ disk is singular on its rim and using (\ref{prolate1}) can be interpreted as the potential associated to a complexified bar of constant density, \cite{letol}. The subsequent members  of the family represents disks of finite density with their maximum on their centers and zero density on their rims. 
\section{Flat Rings}
\begin{figure}
\centering
\epsfig{width=5cm, height=6.5cm, angle=-90,file=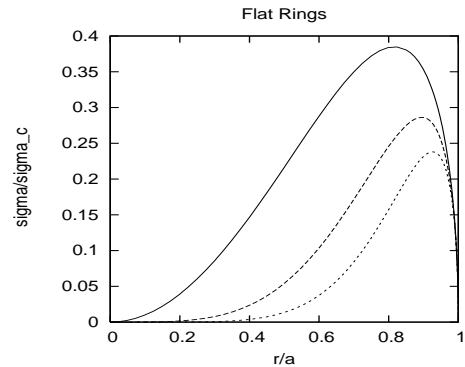}
\caption{The surface density of the first three members of a family disks with a central hole, $\{\sigma^{(n)}_{1r}\}$, i.e., flat rings, the size of the central hole is bigger for larger $n$. }
\label{1ring}
\end{figure}

Let us consider  disks  of the same radius $a$ and decreasing mass,
\be
M^{(n)}=2\pi\sigma_c  a^2/(2n+1), \lb{sigc}
\ee
where  $\sigma_c$ is a constant and will be taken equal for all disks of the Morgan and Morgan family, $n=1,2,3,...$. For these family of disks we have that the corresponding surface density  is
\be
\sigma^{(n)}=\sigma_c (1-p^2)^{\half} (1-p^2)^{n-1}. \lb{sign}
\ee
Now let us consider the following superposition,
\ba 
\sigma^{(n)}_{1r}&=& \sum_{k=0}^{n}C^{n}_k (-1)^{n-k} \sigma^{(n+1-k)},  \\
 &=&\sigma_c (1-p^2)^{\half}p^{2n},\lb{sig1rn}
\ea
where $C^{n}_k=n!/[(n-k)!k!]$. We have that all these superpositions give disks of radius $a$ with zero density on their centers, i.e., disks with a hole in their centers, in other words flat rings. The densities  associated to the three first members of this family of flat rings are presented in Fig. \ref{1ring}. We see a disk with
a hole in the center with  a residual density that is small for larger $n$.
The mass of the three first flat rings are $\pi\sigma_c a^2/16,\;$ $\pi\sigma_c a^2/32,\;$ and $5\pi\sigma_c a^2/256$, respectively.
The potentials associated to these flat  rings can be  found using a superposition
with the same coefficients as the ones used for the densities, e.g., the potential associated to $\sigma_{1r}$ is $V_{1r}= V^{(1)} -V^{(2)}$, etc.

Now let us consider, instead a superposition of disks, a   superposition of  flat  rings,
\be 
\sigma_{2r}^{(n)}=\sigma_{1r}^{(n)}-2b^2\sigma_{1r}^{(n)}+b^4\sigma_{1r}^{(n)},
\ee
where $b$ is a constant such that  $b>a$. This superposition put a gap at $p=1/b$ 
 $(r=a/b)$  on the flat one ring given us a family of two concentric flat rings.
\begin{figure}
\centering
\epsfig{width=5cm, height=6.5cm, angle=-90,file=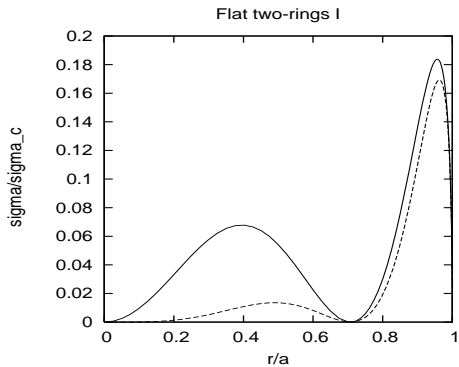}
\caption{The surface density  of the first two members of a family disks with a center hole and a gap, $\sigma_{2r}^{(1)}$ and $\sigma_{2r}^{(2)}$,  i.e., flat  double rings, the size of the central hole is bigger for larger $n$. We take $b^2=2$. }
\lb{2rings}
\end{figure}
\begin{figure}\centering
\epsfig{width=5cm, height=6.5cm, angle=-90,file=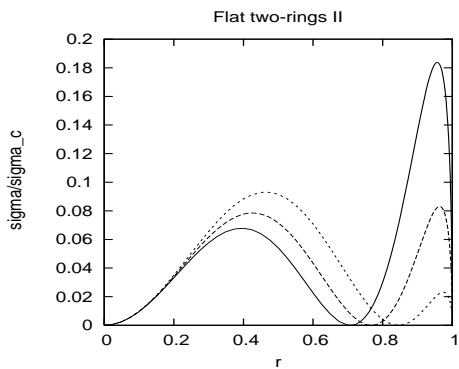}
\caption{The surface density of the first member of the family of two-rings, $\sigma^{(1)}_{2r}$,  for   $b^2=2, 1.7, 1.4$.}
\label{2ringsn}
\end{figure}
In Fig. \ref{2rings}  we present $\sigma_{2r}^{(1)}$ and $\sigma_{2r}^{(2)}$, with $b^2=2$, as before we see that  the  larger the $n$ the larger is the size of the central hole. 
In Fig. \ref{2ringsn} we show  $\sigma_{2r}^{(1)}/ \sigma_c$ for $b^2=2, 1.7, 1.4$.

We can construct several families of two rings using one rings in such  a way that the final density be
\be
 \sigma_{2r}^{(n,m)}=\sigma_c (1-p^2)^{\half} (1-b^2p^2)^{2m} p^{2n} 
\ee
Of course this kind of superposition can be used to have  families of solutions representing $n$ concentric flat  rings with density,
\ba
&&\sigma_{nr}^{(l, \vec{m)}}=\sigma_c (1-p^2)^{\half} (1-b_{(1)}^2p^2)^{2m_1}(1-b_{(2)}^2 p^2)^{m_2} \times  \nn\\
&& \cdot\cdot\cdot \times (1-b_{(n-1)}^2 p^2)^{2m_{n-1}}p^{2l},
\ea
with $\vec m=(m_1, ..., m_{n-1})\in N^{n-1}.$ The positions of the gaps  are  given by $p_i=1/b_i$, with $i=1 ... n-1$, and $b_i>a$.
\section{ Disks with flat rings}
\begin{figure}\centering
\epsfig{width=5cm, height=6.5cm, angle=-90,file=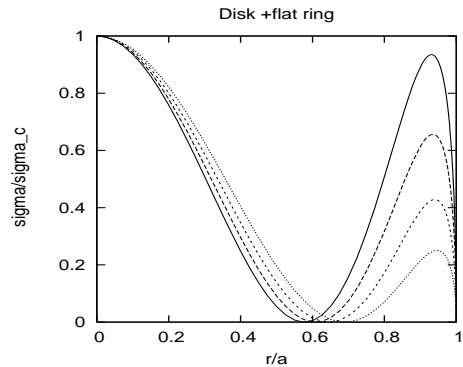}
\caption{The surface density of the first member of the family of disk-one-rings, $ \sigma^{(1)}_{d1r}$, for   $b^2= 3,2.7,2,4,2.1$.    For $b$ small we have a larger central disk.}
\label{diskring}
\end{figure}
By using the same type of superpositions presented in the previous section  we can construct the pair potential-density for disks surrounded by flat rings. Let us consider models of a disk  surrounded by one concentric flat ring. The density
\ba
\sigma^{(1)}_{d1r}&=&\sigma^{(1)}-2b^2\sigma^{(1)}_{(1r)}+ b^4 \sigma^{(2)}_{1r}, \\
& =&\sigma_c (1-p^2)^{\half} (1-b^2p^2)^2,
\ea
represents a disk of  radius $r=a/b$  with a flat ring between  $r=a/b$ and $r=a$. 
In  Fig. \ref{diskring} we present $\sigma^{(1)}_{d1r}/\sigma_c$ for $b^2= 3,2.7,2,4,2.1$
we see that the position of the gap, as well as, the maximum of the flat ring density depends on $b$.

As before, we can build a family of disks with surrounded by one ring with surface density,
\be
\sigma^{(n)}_{d1r}=\sigma_c (1-p^2)^{\half} (1-b^2p^2)^{2n},
\ee
and the general case of a disk surrounded $n$ concentric flat  rings belonging to the above mentioned family,
\ba
\sigma_{dnr}^{(\vec{m})}=\sigma_c (1-p^2)^{\half} (1-b_{(1)}^2p^2)^{2m_1}(1-b_{(2)}^2 p^2)^{2m_2} \times \nn\\
\cdot\cdot\cdot \times  (1-b_{(n)}^2 p^2)^{2m_{n}},
\ea
where now $\vec m=(m_1, ..., m_{n})\in N^{n}.$ 
\section{ Kelvin inverted flat rings}
To construct new family of flat rings we can use the inversion
 theorem of \cite{lord} that in cylindrical coordinates tells us that:
 if the pair potential-density  $V(r,z),\; \sigma(r,z)$ is a solution of Poisson equation the pair,
\ba
&&\hat V(r,z)=V\left[a^2 r/(r^2+z^2), a^2 z/(r^2+z^2)\right],\\
&& \hat\sigma(r,z)=(a/r)^3\sigma\left[a^2 r/(r^2+z^2), a^2 z/(r^2+z^2)\right],
\ea
is also a solution of the same equation, see for instance, \cite{jackson}.
The inversion of the Morgan and Morgan family of disks with  the density (\ref{density2})  gives us,
\be
\hat{\sigma}^{(n)}(r)=(2n+1)M^{(n)}/(2\pi a^2q^3)(1-q^2)^{n-\half},
\lb{deninv1}
\ee
where $q=a/r$ and $n\geq 1$.  
Note that that the Kelvin inverted Morgan and Morgan   disks can be considered as representing flat rings, we have a disk with a hole of radius $a$. The density of  these  flat rings decays in a similar way as  the Plummer-Kuzmin  disks, i.e, as $1/r^3$,  not too fast.

The mass of the Kelvin inverted Morgan and Morgan disks is given by,
\be
\hat{M}^{(n)}=(2n+1)!!(2n)!! M^{(n)}/(2n)!!.
\label{massinv}
\ee
The Kelvin inverted one-rings (\ref{sig1rn}) gives other family of one-rings with density
\be
\hat{\sigma}^{(n)}_{1r}= \sigma_c (1-q^2)^{\half}q^{2n+3}. \lb{sigin1rn}
\ee
Note that the density  decays as   $\hat{\sigma}^{(n)}_{1r}\sim 1/r^{2n+3}$ for this family of flat rings. 
\begin{figure}
\epsfig{width=5cm, height=6.5cm, angle=-90,file=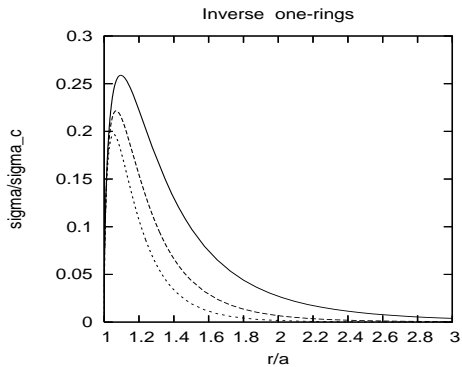}
\caption{The surface density of the first three member of the family
 of inverse one-rings, $\{ \hat{\sigma}^{(n)}_{1r}\}$. We have a central hole of radius $r=a$. Even though the rings extent to infinity we see a  clear cutoff.}
\label{kel1}
\end{figure}
In Fig. \ref{kel1} we plot the inverted density of the first three members of the one-flat-ring family. Note the central hole of radius $r=a$. Due to the fast decay rate with the distance we have that on can put a clear cutoff and to consider this flat rings as finite, Eq. (\ref{sigin1rn}). 
We can also 
invert the families of two and several flat rings discussed in Sec. 3, we will also have new families of rings that decay with the distance very fast.
\section{Stability of circular orbits}
In this section we study the stability under radial perturbations of circular orbits concentric to the flat rings. By assuming that flat rings are made of particles moving along circular orbits the study of the stability of these  orbits can be considered as an order zero test of stability of the flat rings. In this case the collective behavior of the particles of the ring is not taken into account. The epicyclic frequency of the perturbed orbit is, \cite{BT},
\be
\kappa^2= a^2\left(\frac{\partial^2 V}{\partial p^2}+\frac{3}{p}\frac{\partial V}{\partial p} \right). 
\label{kappa}
\ee

The  epicyclic frequencies of rings  with densities $\sigma^{(1)}_{1r}, \sigma^{(2)}_{1r}$ and  their associated Kelvin inverted flat single rings  with densities $\hat{\sigma}^{(1)}_{1r}, \hat{\sigma}^{(2)}_{1r}$ are: 
\ba
\kappa ^{(1)}_{1r}&=&\pi \left[ a^3\sigma_c (27p^2 -8)/8\right]^\half ,\;\;(0\leq p \leq 1), \\
\kappa ^{(2)}_{1r}&=&\frac{\pi}{4}\left[ a^3\sigma_c (105p^4-81p^2+20)\right]^\half ,\;\; (0\leq p \leq 1),\\
\hat{\kappa}^{(1)}_{1r}&=&3\frac{\pi}{p^3} \left[ a^3\sigma_c/8)\right]^\half, \;\; (p\geq 1), \\
\hat{\kappa}^{(1)}_{1r}&=&\frac{\pi}{p^4} \left[ 3a^3\sigma_c (-18p^2+35)/32\right]^\half,\;\; ( p\geq 1). 
\ea
The one-ring with density $\sigma^{(1)}_{1r}$ is not completely stable, we have stability only for $r>(2\sqrt{2}/3)a$. The flat rings with densities, $\sigma^{(2)}_{1r}$, and  $\hat{\sigma}^{(1)}_{1r}$ are stable in all their extensions. We have that only the inner  part, $ 1<(r/a)<
\sqrt{35/18}$,  of the flat ring with density $\hat{\sigma}^{(2)}_{1r}$ is stable.
\section{Discussion}
We  presented several families of pairs potential-density that can be used to represent one or several concentric rings, also a disk surrounded by 
flat rings. We have shown in some detail the superposition of surface densities to build the families indicating that the same superposition is valid to build the associated potential. 

We have considered only the density as representing the physics inside the different structures studied along this work, in principle, one can find the other thermodynamics variables by solving the Fokker-Plank equation. In a recent work we presented a method for solving this equation for a disk, \cite{maxlet}, that with simple modifications can be used to find a distribution function for these structures. 

With our very first approach to the stability of flat  rings presented in the penultimate section, we find stable configurations, as well as, partially stable ones.  The structure of one-ring standing alone  is not very realist. For instance in the  case  of planetary flat rings we have a planet  located  in center of the ring  that will contribute to  stabilize the structure. To be more specific let us consider a  planet with mass $M$  surrounded by a flat ring. Let   $\kappa_{r}^2$  be the square of the epicyclic frequency associated to the ring alone. Now the square of the  epicyclic frequency of a particle orbiting the system planet-ring  is  
$\kappa_{pr}^2= \kappa_{r}^2 +GMa^3/p^3$. Since the contribution of the planet to the epicyclic frequency of the particles of the disk  is a positive quantity we have that the 
planet tends to stabilize the rings. To perform a study of the stability of the structures 
above described taken into account the collective behavior of the particles  we need to  know the  fluid dynamics variables that in principle can be found solving the Fokker-Planck equation using the method already mentioned.
\section*{Acknowledgments}
I  thank CNPq and FAPESP for financial support.

\end{document}